\documentclass[preprint,12pt,nopreprintline]{elsarticle}

\usepackage{amssymb}
\usepackage{amsmath}
\usepackage{booktabs}
\usepackage{graphicx}
\usepackage{subcaption}
\usepackage{hyperref}
\usepackage{color}

\journal{Computational Materials Science}

\begin{document}
\begin{frontmatter}

\title{Benchmarking of Fast and Interpretable UF Machine Learning Potentials}

\author[phys,qtp]{Pawan Prakash}

\author[qtp,mse]{Sam Dong}

\author[qtp,mse]{Richard Hennig\corref{cor1}}
\ead{rhennig@ufl.edu}

\affiliation[phys]{organization={Department of Physics, University of Florida},
                  city={Gainesville},
                  state={FL},
                  postcode={32611},
                  country={USA}}
                  
\affiliation[qtp]{organization={Quantum Theory Project, University of Florida},
                  city={Gainesville},
                  state={FL},
                  postcode={32611},
                  country={USA}}

\affiliation[mse]{organization={Department of Materials Science and Engineering, University of Florida},
                  city={Gainesville},
                  state={FL},
                  postcode={32611},
                  country={USA}}

\cortext[cor1]{Corresponding author}

\begin{abstract}
  Machine learning interatomic potentials (MLIPs) have emerged as a powerful alternative to density functional theory (DFT) for molecular dynamics simulations, offering near-DFT accuracy at a fraction of the computational cost. However, many state-of-the-art MLIPs remain computationally demanding and act as black boxes, limiting physical interpretability. In this work, we evaluate the ultra-fast force field (UF$^3$) potential, which employs linear regression with cubic B-spline basis to represent effective two- and three-body interactions. We show that UF$^3$ displays accuracy comparable to established models such as GAP, MTP, NNP (Behler Parrinello), and qSNAP MLIPs. We further investigate the transferability of UF$^3$ by computing melting points for six elemental systems with potentials fitted without any solid--liquid interface configurations or explicit thermodynamic information about melting. The model reproduces experimental melting points within $\sim$6\% for simple metals (Ni, Cu, Li), but substantially underestimates them for Mo and Si and fails to yield a stable potential for Ge, reflecting the limitations of a fixed expansion truncated at the three-body term for systems with strong angular or covalent bonding. We further illustrate how UF$^3$'s spline-based formulation allows direct visualization of the learned interactions, enabling identification of unphysical behavior that black-box approaches often obscure.
\end{abstract}

\begin{keyword}
machine learning interatomic potentials \sep UF$^3$ \sep computational materials science \sep molecular dynamics \sep interpretability
\end{keyword}
\end{frontmatter}

\section{Introduction}

Molecular dynamics (MD) simulations play a pivotal role in understanding and predicting the behavior of materials at atomic and molecular scales \cite{THOMPSON2022108171}. Density functional theory (DFT) \cite{kohn1964inhomogeneous, kohn1965self}, a quantum mechanical modeling method, offers detailed insights and precise predictions about these systems. However, the high computational costs and $O(N)-O(N^3)$ scaling associated with DFT make it challenging for use in extensive MD simulations spanning long durations \cite{mattsson2005designing}. Classical interatomic potentials \cite{jones1924determinationI, JELennard-Jones_1931} offer computational efficiency, but lack DFT's accuracy \cite{doi:10.1021/ja00051a040, JGodet_2003}. The advent of machine learning interatomic potentials (MLIPs) has provided an effective alternative, combining the efficiency of classical potentials with near-DFT accuracy in energy, force, and property predictions \cite{PhysRevLett.98.146401, PhysRevLett.104.136403, doi:10.1137/15M1054183, THOMPSON2015316}. MLIPs also play a significant role in accelerating materials discovery \cite{prakash2025guidediff, beenet25}. However, a challenge arises with many state-of-the-art MLIPs: while they are more efficient than DFT, they can still be computationally demanding, and many act as a quasi black box, limiting the potential's physical interpretability.

In the landscape of machine learning potentials, several have emerged as prominent contenders in accurately modeling atomic systems. These include the neural network potentials (NNP) \cite{PhysRevLett.98.146401}, Gaussian approximation potential (GAP) \cite{PhysRevLett.104.136403}, moment tensor potential (MTP) \cite{doi:10.1137/15M1054183}, spectral neighbor analysis potentials (SNAP) \cite{THOMPSON2015316}, and its variant, quadratic SNAP (qSNAP) \cite{10.1063/1.5017641}. Central to the efficacy of these potentials is the concept of `fingerprints' or `descriptors'. These are intricate representations of atomic environments with flexible functional form, which are subsequently passed through an embedding function to yield the desired predictions. % For readers desiring a comprehensive understanding of each method's details, the provided references offer an in-depth exploration.

This study builds on previous work comparing machine learning potentials in energy and force prediction accuracy using a consistent dataset of single element systems\cite{doi:10.1021/acs.jpca.9b08723}. The primary focus of this study is the ultra-fast force field machine learning interatomic potential (UF$^3$ - MLIP) \cite{xie2023ultrafast}, where we benchmark UF$^3$'s predicting capabilities against prevailing MLIPs.
Beyond accuracy benchmarks, we examine the transferability of UF$^3$ to melting behavior using potentials fitted without any solid--liquid interface configurations or explicit thermodynamic information about melting. We find that transferability is element-dependent: it succeeds for simple metallic bonding (Ni, Cu, Li) but degrades for systems with strong directional or covalent bonding (Mo, Si, Ge), reflecting the intrinsic limitations of a fixed expansion truncated at the three-body term. We additionally illustrate UF$^3$'s inherent interpretability advantage: its spline basis permits direct visualization of effective interatomic interactions, providing a practical tool for diagnosing unphysical behavior in the fitted potential.

\section{Methods}

\subsection{UF$^3$ Theory}

The UF$^3$-MLIP is formulated as a function of atomic positions $R_i$ and species $\sigma_i$. It approximates the many-body expansion of the energy of atomic systems. By truncating its expansion to the three-body term, UF$^3$ strikes a favorable compromise between computational efficiency and predictive accuracy \cite{xie2023ultrafast}:
\begin{equation}
    E = \sum_{i,j} V_2(r_{ij}) + \sum_{i,j,k} V_3(r_{ij}, r_{ik}, r_{jk})
    \label{e23}
\end{equation}
where indices $i, j, k$ of the summation run over all atoms, $V_2(r_{ij})$ is the two-body contribution and $V_3(r_{ij}, r_{ik}, r_{jk})$ is the three-body contribution.

Using a spline basis with compact support, UF$^3$ represents effective two- and three-body interactions. It employs cubic B-splines and tensor product splines for smooth and efficient representation:
\begin{equation}
    V_2(r_{ij}) = \sum^K_{n=0} c_n B_n(r_{ij})
\end{equation}
\begin{equation}
    V_3(r_{ij}, r_{ik}, r_{jk}) = \sum^{K_l}_{l=0} \sum^{K_m}_{m=0} \sum^{K_n}_{n=0} c_{lmn} B_l(r_{ij}) B_m(r_{ik}) B_n(r_{jk})
\end{equation}

UF$^3$'s training uses regularized linear regression to optimize the coefficients $c_n$ and $c_{lmn}$ of the two-and three-body terms, such that energies and forces are fit to data from ab initio calculations. The loss function incorporates Tikhonov regularization. The ridge regularizer $\lambda_1$ imposes a penalty on large coefficients to prevent overfitting and mitigate coefficient fluctuation in response to change in data. Meanwhile, the curvature regularizer $\lambda_2$ ensures the smoothness of the learned potential. The loss function for two-body is
\begin{equation}
\begin{split}
    L &= \frac{\kappa}{\sigma^2_\mathcal{E}|\mathcal{E}|}\sum_{s\in S} (E(s)-\mathcal{E}_s)^2 + \frac{1-\kappa}{\sigma^2_\mathcal{F}|\mathcal{F}|}\sum_{s\in S} (-\nabla E(s)-\mathcal{F}_s)^2 \\
    &\phantom{=} + \lambda_1\sum^K_n c_n^2 + \lambda_2\sum^K_n(c_n - 2c_{n+1} + c_{n+2})^2
    \label{loss}
\end{split}
\end{equation}
where $\kappa \in [0,1]$ is the effective weight of energy/force residual, $E(s)$ is the predicted energy, and $|\mathcal{F}|$ and $|\mathcal{E}|$ denote the number of forces and energy observation in training set, respectively.

A standout attribute of the UF$^3$ framework is its interpretability. Given its reliance on tangible physical parameters like inter-atomic distances, it paves the way for straightforward visualization of its model. This allows for easy identification of any anomalies or unphysical behaviors in the fitted potential.

\subsection{Data and Hyperparameter Optimization}

The UF$^3$ potential adopts a two-tiered hyperparameter system. The `outer' hyperparameters in featurization guide both initialization and the learning of atomic environments, while the `inner' hyperparameters of the loss function direct the least square optimization. Specifically, the outer hyperparameters set the maximum and minimum cutoff distances between atoms and determine the number of basis splines for each spline. In contrast, the inner hyperparameters address aspects like the effective weight of the energy/force residual in the loss function and the regularization parameters, namely the two- and three-body ridge and curvature regularizers. It's important to underline that adjustments to the outer hyperparameters necessitate the generation of new representation vectors, a process that is both time-intensive and computationally demanding, making it the most resource-intensive step in the UF$^3$ framework. %This distinction of hyperparameters into tiers is based on their computational complexity.

Table~\ref{tab:hyperparams} summarizes the optimized UF$^3$ hyperparameters used for each elemental system in this study. We use the pre-defined 90:10 train/test splits published by Ong et al.~\cite{doi:10.1021/acs.jpca.9b08723}, which makes our test errors directly comparable to the published errors of GAP, MTP, NNP, SNAP, and qSNAP. The inner hyperparameters were selected by grid search with 5-fold cross-validation on the training split only, over per-element grids of 300--7{,}776 points spanning ridge and curvature regularization strengths from $10^{-3}$ to $10^{-8}$ (plus zero) and, where scanned, energy weights $\kappa \in \{0.3, 0.5, 0.7\}$; candidate hyperparameter sets were ranked by a normalized cross-validation cost weighting energy errors at 0.7 and force errors at 0.3, and the final model for each element was refit on the full training split with the selected values; all reported errors are evaluated on the held-out test split. Because the UF$^3$ loss function (Eq.~\ref{loss}) is strongly convex and is solved directly via LU decomposition~\cite{xie2023ultrafast}, the fit is deterministic: identical training data and hyperparameters yield identical spline coefficients, so reproducibility of the fits is fully captured by the values below.

\begin{table}[htbp]
\centering
\caption{Optimized UF$^3$ hyperparameters and training settings per element. $r_{\rm cut}^{(2)}$ is the outer cutoff of the two-body spline (with $r_{\rm min}^{(2)} = 0.001$~\AA\ for all elements); $r_{\rm min}^{(3)}$ and $r_{\rm cut}^{(3)}$ delimit the equivalent $r_{ij}$ and $r_{ik}$ axes of the three-body spline (the $r_{jk}$ axis extends to $2\cdot r_{\rm cut}^{(3)}$ by triangle inequality). $N^{(2)}$ and $N^{(3)}$ are the numbers of cubic B-spline basis functions, related to the user-specified resolution (number of intervals between unique knot positions) by $N = \mathrm{resolution} + 3$. $\lambda_1$ and $\lambda_2$ are the two-/three-body ridge and curvature regularizers (a ridge of $10^{-8}$ is additionally applied to the one-body offset term); $\kappa$ is the energy weight of the loss function (Eq.~\ref{loss}) used in the fit --- all six elements selected $\kappa = 0.3$. Train/test gives the number of structures in the pre-defined 90:10 splits provided by Ong et al.~\cite{doi:10.1021/acs.jpca.9b08723}.}
\label{tab:hyperparams}
\footnotesize
\setlength{\tabcolsep}{3pt}
\resizebox{\textwidth}{!}{%
\begin{tabular}{lccccccccc}
\toprule
\textbf{Element} & $r_{\rm cut}^{(2)}$ (\AA) & $N^{(2)}$ & $r_{\rm min}^{(3)}$ (\AA) & $r_{\rm cut}^{(3)}$ (\AA) & $N^{(3)}$ & $\lambda_1$ (2b/3b) & $\lambda_2$ (2b/3b) & $\kappa$ & Train/test \\
\midrule
Ni & 4.00 & 19 & 1.20 & 3.70 & $12{\times}12{\times}18$ & $0\,/\,10^{-8}$ & $10^{-5}\,/\,10^{-8}$ & 0.3 & 263/31 \\
Cu & 4.10 & 18 & 1.50 & 3.70 & $11{\times}11{\times}19$ & $0\,/\,10^{-8}$ & $10^{-5}\,/\,10^{-8}$ & 0.3 & 262/31 \\
Li & 5.10 & 19 & 1.10 & 4.60 & $12{\times}12{\times}18$ & $0\,/\,10^{-4}$ & $10^{-5}\,/\,10^{-4}$ & 0.3 & 241/29 \\
Mo & 5.20 & 20 & 1.50 & 4.70 & $11{\times}11{\times}19$ & $0\,/\,10^{-5}$ & $10^{-8}\,/\,10^{-8}$ & 0.3 & 194/23 \\
Si & 5.30 & 19 & 1.42 & 4.70 & $11{\times}11{\times}19$ & $0\,/\,10^{-4}$ & $10^{-6}\,/\,10^{-8}$ & 0.3 & 214/25 \\
Ge & 5.40 & 19 & 1.30 & 5.00 & $13{\times}13{\times}20$ & $0\,/\,10^{-5}$ & $10^{-5}\,/\,10^{-5}$ & 0.3 & 228/25 \\
\bottomrule
\end{tabular}%
}%

\end{table}

\section{Results}
Although recent advances in MLIPs are commendable, a meticulous and systematic evaluation of MLIPs on standardized datasets remains largely elusive. It is in this context that the efforts of Ong et al. \cite{doi:10.1021/acs.jpca.9b08723} stand out; they put forth a comprehensive dataset for MLIP comparison, which encompasses a diverse range of atomic local environments associated with six elements: Ni, Cu, Li, Mo, Si, and Ge. Selected for their diverse crystal structures and chemistries, the dataset covers ground-states, strained structures, ab initio MD simulations of bulk supercells at elevated temperatures, and vacancy calculations. The AIMD snapshots span 300~K up to approximately twice the melting temperature of each element, and the snapshots from above the melting temperature sample disordered, liquid-like local environments \cite{doi:10.1021/acs.jpca.9b08723}. We use the pre-defined datasets of Ong et al.\ in full; no structure group or temperature was excluded for any element.
We will be using the dataset of Ong et al. \cite{doi:10.1021/acs.jpca.9b08723} for our benchmarking.

\subsection{Energy and Forces}
\begin{figure}[ht]
    \centering
    \begin{subfigure}{0.45\textwidth}
        \centering
        \includegraphics[width=\linewidth]{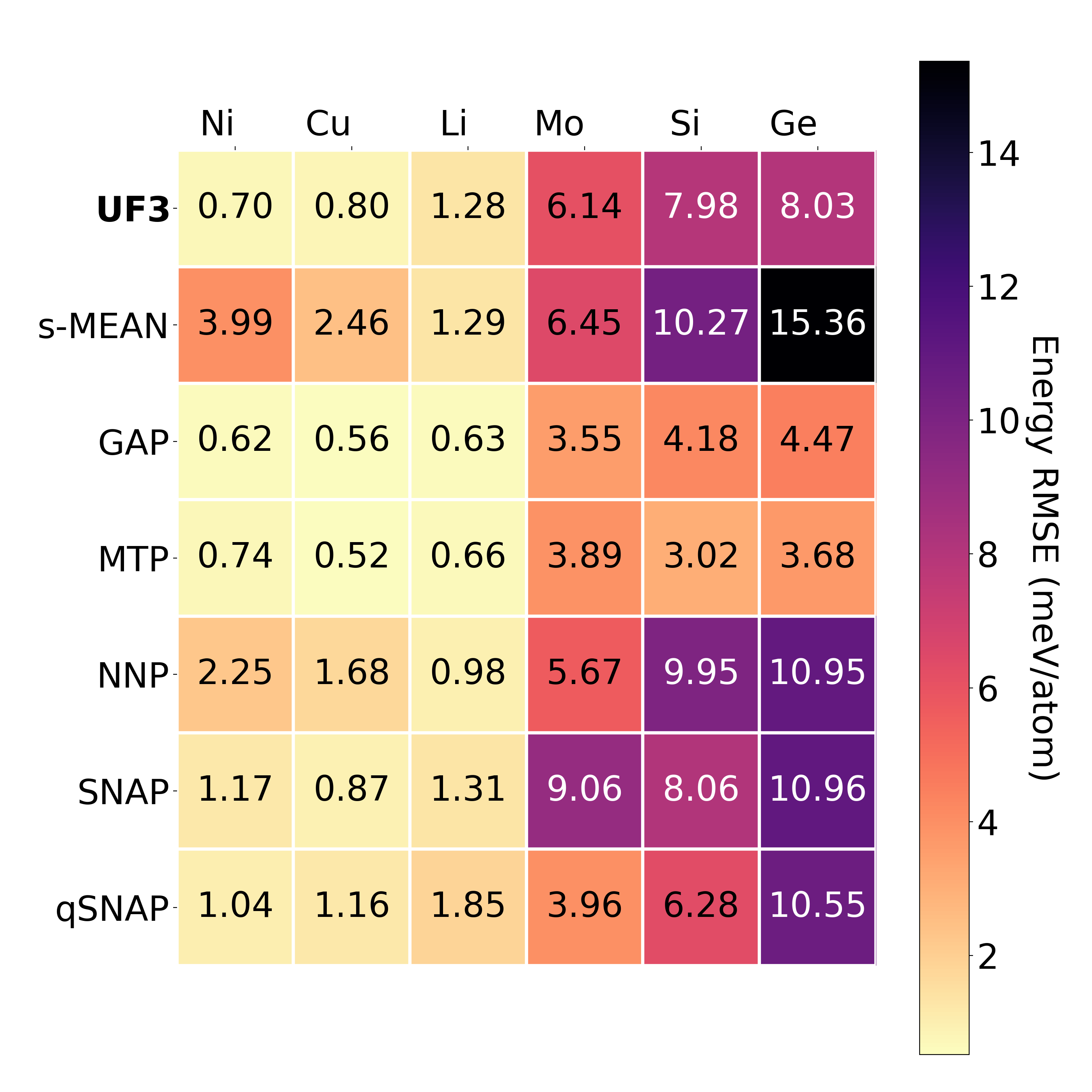}
    \end{subfigure}%
    ~ % spacing between the subfigures
    \begin{subfigure}{0.45\textwidth}
        \centering
        \includegraphics[width=\linewidth]{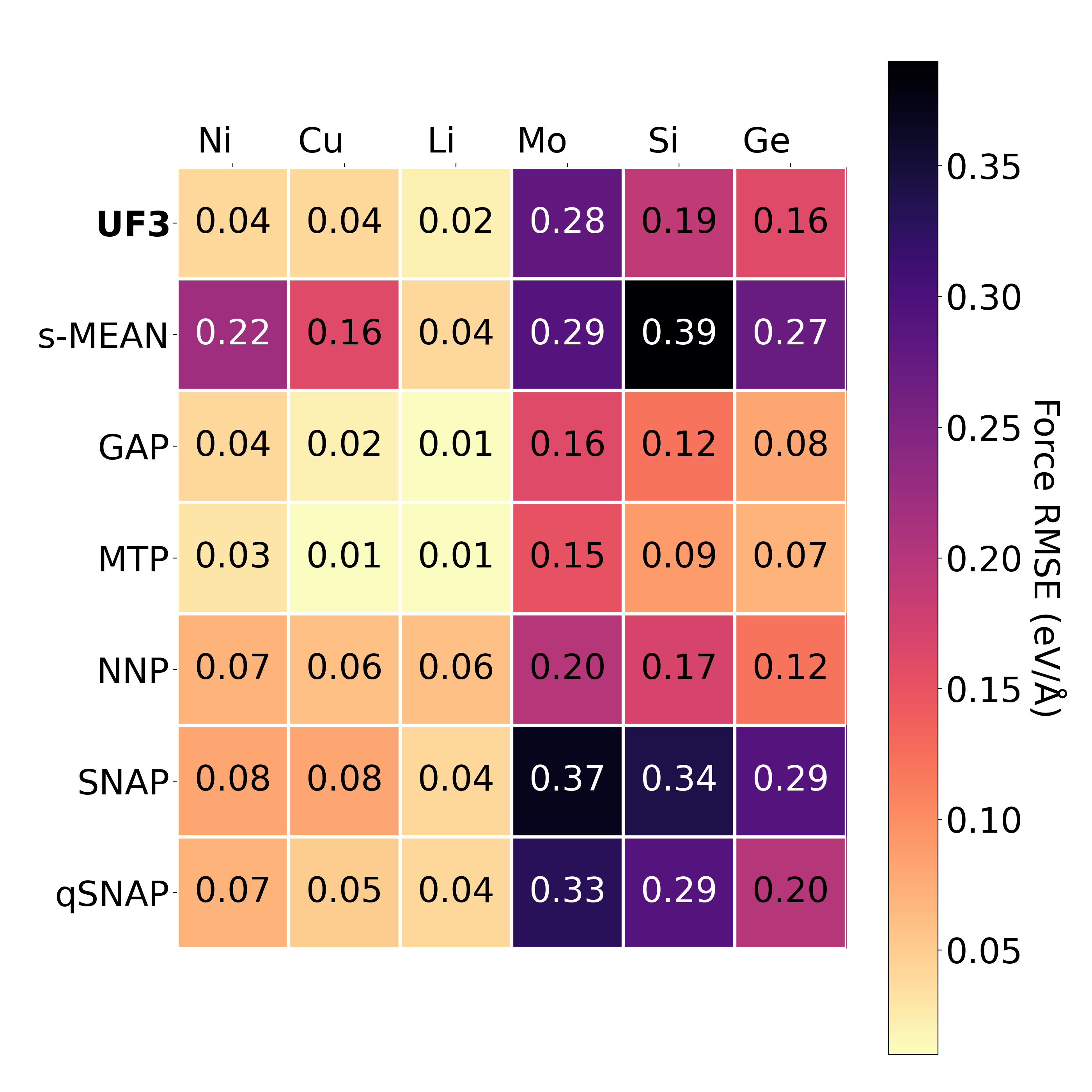}
    \end{subfigure}
    \caption{Energy errors (left) and Force errors (right) for UF$^3$ compared to other MLIPs. UF$^3$ exhibits comparable accuracy to GAP, MTP, and qSNAP for FCC crystals, highlighting its efficiency. However, its precision is slightly lower for more open structures like diamond and BCC as well as others, but it still ranks third or fourth in terms of accuracy. }
    \label{fig:energy_force}
\end{figure}

Ong et al. performed a detailed comparison of state-of-the-art MLIPs in energy and force prediction for elemental metals \cite{doi:10.1021/acs.jpca.9b08723}. To ensure a consistent comparison, we optimized our cutoff radius hyperparameter within similar bounds specified in Ong et al. Following the inner-hyperparameter optimization, the resultant energy and force errors are depicted in Fig. \ref{fig:energy_force}. These figures underscore that UF$^3$ offers accuracy on par with GAP, MTP, and qSNAP MLIPs for face-centered cubic (FCC) crystals, while significantly reducing computational cost \cite{xie2023ultrafast}. The computational cost of UF$^3$ was characterized in detail for bcc tungsten in the genesis paper~\cite{xie2023ultrafast}: UF$_{2,3}$ was reported at $\sim$10.5$\times$ the cost of a Lennard-Jones (LJ) pair potential, compared with MTP at $\sim$46$\times$, qSNAP at $\sim$145$\times$, SNAP at $\sim$443$\times$, and GAP at $\sim$3070$\times$ LJ, corresponding to speed-ups ranging from a factor of $\sim$4 (MTP) to three orders of magnitude (GAP) at similar accuracy. Because UF$^3$'s cost scales with the number of basis functions and the cutoff radius --- both of which are comparable across the six elements considered here to those used by Xie et al.\ for tungsten --- this cost advantage is expected to transfer. Per-element cost benchmarks for the comparison MLIPs are reported in Ong et al.~\cite{doi:10.1021/acs.jpca.9b08723}. For the potentials developed in this work, we measure evaluation costs of 21.6--25.9~$\mu$s/(atom$\cdot$step) for Ni, Cu, and Li and 55.6--65.6~$\mu$s/(atom$\cdot$step) for Mo, Si, and Ge in LAMMPS on a single CPU core (AMD EPYC 7702) for bulk supercells of $\sim$11{,}000 atoms; the higher cost of the second group reflects their larger three-body cutoffs and denser neighbor environments, since the three-body term loops over pairs of neighbors. On the identical core and with the identical executable, these costs correspond to 15--18$\times$ that of EAM for the fcc metals and 34--57$\times$ that of the Stillinger--Weber and Tersoff potentials for the diamond-structure semiconductors --- placing UF$^3$ within one to two orders of magnitude of classical empirical potentials, a regime that MTP, SNAP, qSNAP, and GAP exceed by further factors of $\sim$4 to $\sim$300 in the controlled same-hardware comparison of Xie et al.~\cite{xie2023ultrafast}. In more open geometries like diamond and body-centered cubic (BCC), all MLIPs, including UF$^3$, showed diminished accuracy, with UF$^3$ ranking third among all MLIPs. Despite this, UF$^3$ and other MLIPs outperformed spline-based empirical potentials such as s-MEAN \cite{VITA2021110752}.

\begin{figure}[htbp]%!hb]
    \centering
    \begin{subfigure}{0.4\textwidth}
        \centering
        \includegraphics[width=\linewidth]{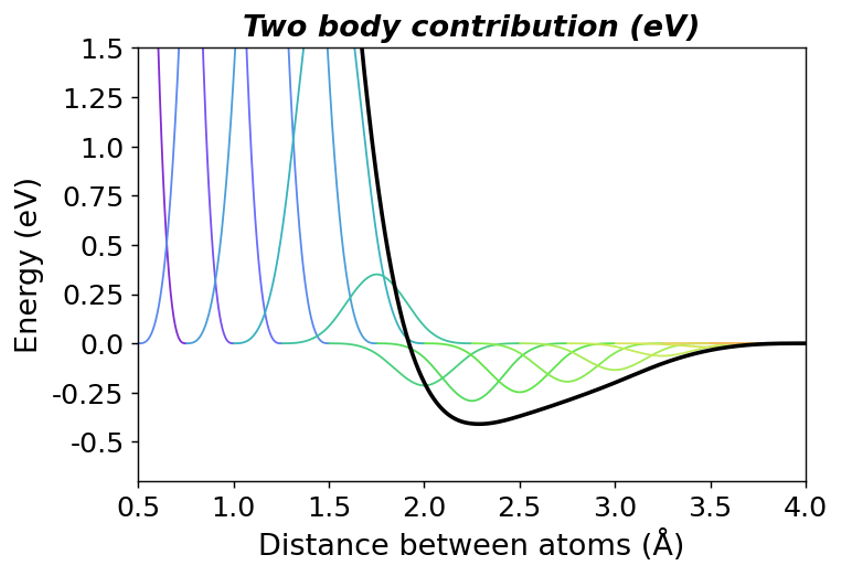}
        \caption{Two-body contribution}
        \label{fig:vis1}
    \end{subfigure}%
    ~ % spacing between the subfigures
    \begin{subfigure}{0.5\textwidth}
        \centering
        \includegraphics[width=\linewidth]{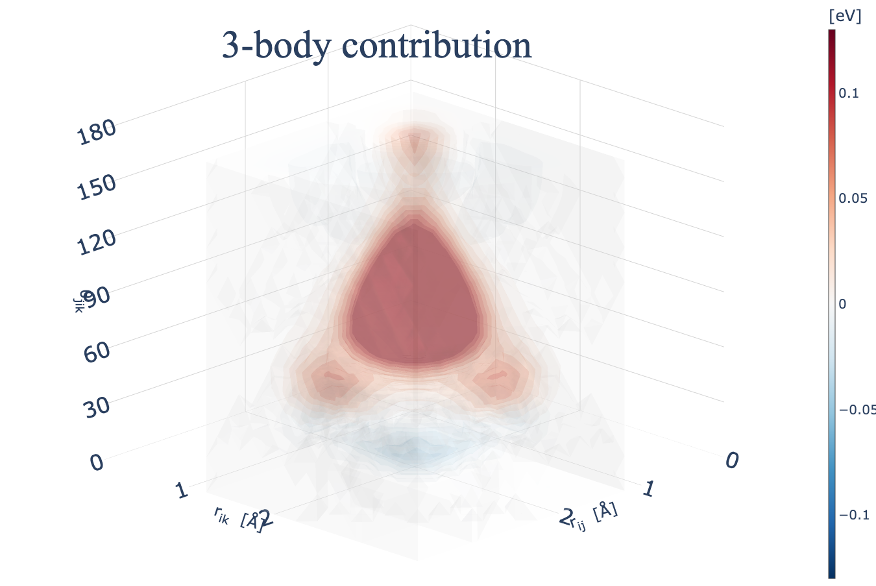}
        \caption{Three-body contribution}
        \label{fig:vis2}
    \end{subfigure}
    ~ ~ % spacing between the subfigures
    \begin{subfigure}{0.9\textwidth}
        \centering
        \includegraphics[width=\linewidth]{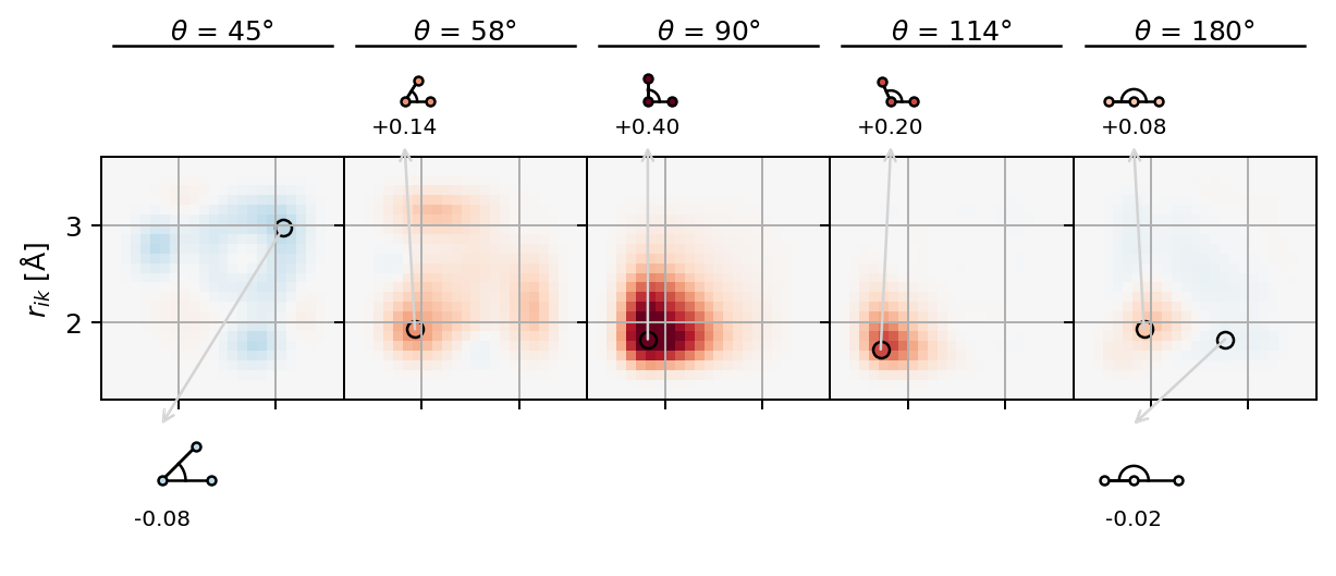}
        \caption{Sliced view of three-body contributions at certain angles.}
        \label{fig:vis3}
    \end{subfigure}
    \caption{ Visualization of the learned UF$^3$ potential for nickel. (a)~Two-body contribution $V_2(r_{ij})$ (black) and the underlying B-spline basis (colored), showing a steep short-range repulsion, a well minimum of $-0.41$~eV at 2.3~\AA\ slightly inside the FCC nearest-neighbor distance (2.49~\AA), and smooth decay to the cutoff. (b)~Three-dimensional rendering of the three-body contribution $V_3(r_{ij},r_{ik},\theta_{jik})$, with red and blue volumes marking unfavorable and favored angular configurations. (c)~Slices of $V_3$ at fixed angles: strongest unfavorable contribution at $\theta=90^\circ$ ($r_{jk}\approx2$~\AA), with weaker features at $58^\circ$ and $114^\circ$. All learned features --- well depth, equilibrium distance, angular preferences, and unphysical artifacts --- are directly readable from the plots, providing a transparency not available for descriptor-based MLIPs.}
    \label{fig:visualization}
\end{figure}

Fig.~\ref{fig:visualization} illustrates the interpretability of UF$^3$ for nickel: because the potential is an explicit sum of two- and three-body splines~\cite{xie2023ultrafast}, the learned functions can be plotted directly as functions of physical distances and angles, unlike descriptor-based models such as GAP, MTP, NNP, or SNAP.
Panel~(a) shows the fitted two-body potential $V_2(r_{ij})$ and the underlying B-spline basis (colored curves) whose weighted sum yields the black envelope. The curve exhibits the characteristic features of a metallic pair interaction: a steep short-range repulsion below $\sim$2~\AA, a well minimum of $-0.41$~eV at 2.3~\AA, slightly inside the FCC nearest-neighbor distance (2.49~\AA\ $\approx a/\sqrt{2}$) --- as expected, since the equilibrium spacing reflects the balance of the pair attraction with contributions from more distant coordination shells and the three-body term --- and a smooth monotonic decay to the cutoff. The absence of spurious oscillations or non-monotonic regions beyond the minimum indicates a well-regularized fit; conversely, such features would directly signal an overfit potential.
Panel~(b) renders the full three-body contribution $V_3(r_{ij},r_{ik},\theta_{jik})$, with red (positive) and blue (negative) volumes marking unfavorable and favored angular configurations. Panel~(c) quantifies this by slicing at fixed angles: the dominant unfavorable feature is at $\theta=90^\circ$ ($\sim+0.40$~eV, $r_{jk}\approx2$~\AA), with weaker features at $58^\circ$ and $114^\circ$ ($+0.14$ and $+0.20$~eV) and small negative contributions at $45^\circ$ and $180^\circ$. This mild, smooth angular dependence is consistent with the close-packed geometry of FCC nickel and contrasts with the sharp tetrahedral preference required for covalent diamond-cubic systems (Section~3.3). All learned features --- equilibrium distance, well depth, angular preferences, and the presence or absence of unphysical artifacts --- are directly readable from the plotted functions, providing a diagnostic tool not available for descriptor-based MLIPs.

Read together, the features in Fig.~\ref{fig:visualization} also constitute a diagnostic check on the fitted potential. A $V_2$ that is non-monotonic at short range, fails to decay at $r_{\rm cut}$, or has its minimum displaced from the expected nearest-neighbor distance signals insufficient curvature regularization ($\lambda_2$), an under-extended cutoff, or insufficient knot density, respectively; a $V_3$ with jagged angular features or preferences at unphysical angles signals an over-flexible three-body basis or under-regularization. Knot density should track the radial distribution function of the training data~\cite{xie2023ultrafast}; because UF$^3$ is linear in localized basis functions, every such feature is attributable to specific coefficients, and the data coverage of each basis function is available directly from the featurization, so poorly constrained regions of configuration space can be identified explicitly. This visual screening also exposes limitations that no numerical $E/F$ metric can detect: a \emph{fixed} two- plus three-body expansion lacks the environment-dependent (bond-order) rescaling of the interactions that transferable potentials for tetrahedral semiconductors rely on~\cite{tersoff1988new,stillinger1985computer}, consistent with our experience for Si and Ge in Section~3.3.

Applying this screening across all six elements (Fig.~\ref{fig:v2all} in \ref{app:v2all}) makes the diagnostics concrete: the elements with accurate melting points in Section~3.3 (Ni, Cu, Li) exhibit smooth single-minimum pair terms; Mo develops a double-well structure mirroring the bcc first- and second-neighbor shells; and Ge exhibits a spurious repulsive barrier between its coordination shells, larger in magnitude than its own binding minimum, despite competitive test errors (8.0~meV/atom, 0.16~eV/\AA). The Ge artifact, co-located with the sparsely sampled region between coordination shells, anticipates the instability of the Ge melting simulations: the visualization identifies a failure mode that energy and force metrics miss.

\subsection{Property Predictions}

\begin{figure}[htbp]
\centering
\includegraphics[width=0.9\textwidth]{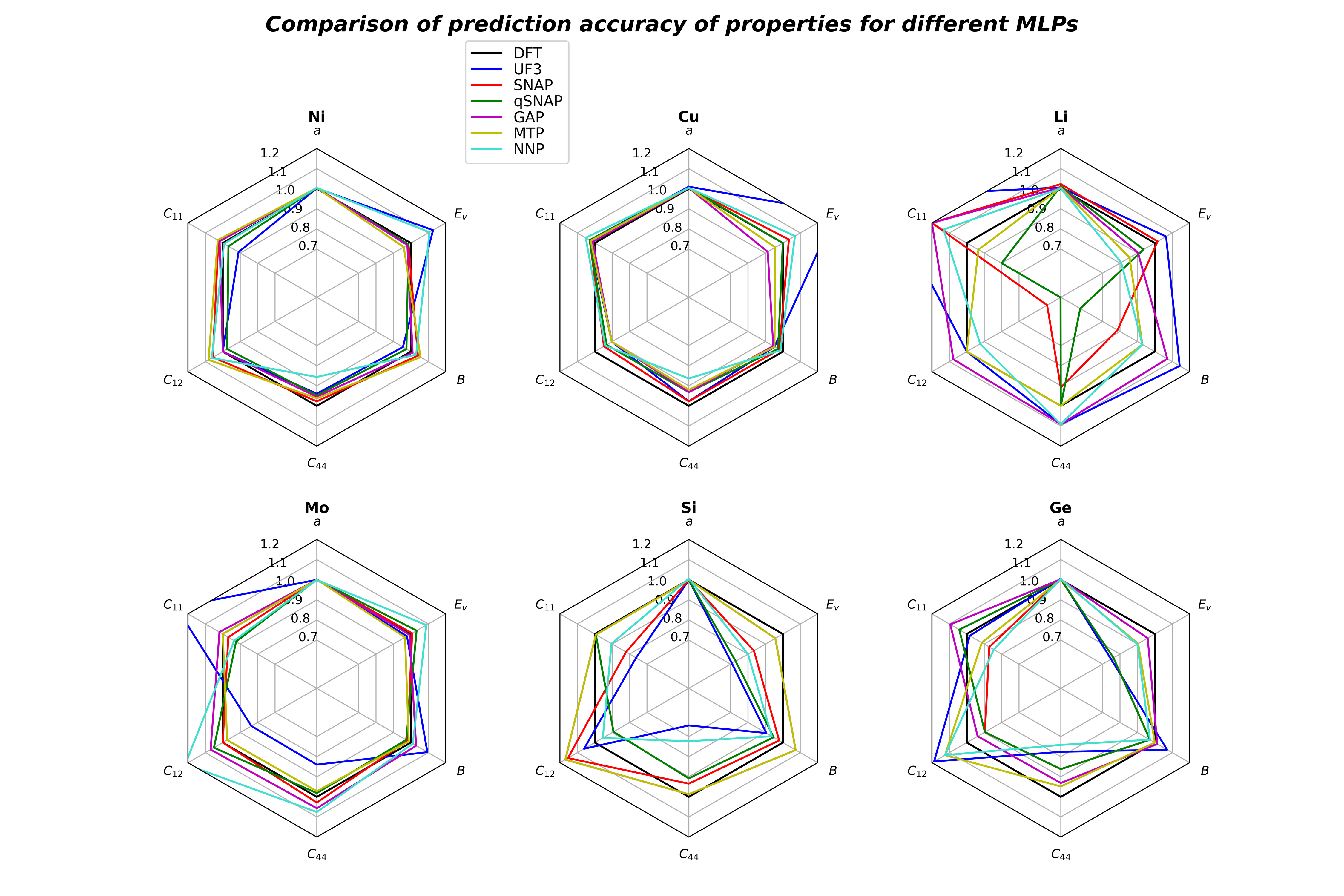}
\caption{Spider plots comparing the predictive capabilities of various MLIPs with DFT benchmarks for Ni, Cu, and Li. Properties shown include cubic elastic constants $C_{11}, C_{12}, C_{44}$, bulk modulus ($B$), vacancy formation energy ($E_v$), and lattice constant ($a$). Values are normalized to DFT results.}
\label{fig:spider}
\end{figure}

We evaluated property predictions using UF$^3$ models optimized for energy and force, comparing them with DFT data and other advanced MLIPs. These comparisons, illustrated in spider plots (Fig. \ref{fig:spider}), show UF$^3$'s high predictive accuracy for properties like cubic elastic constants, bulk modulus, and lattice constants, particularly for Ni and Cu. Discrepancies notably arise in predicting properties for Li, where UF$^3$ and other MLIPs diverge from DFT results, partly due to the small magnitude of Li's properties. This variation in performance across elements highlights the critical role of having diverse and representative training data.
Although the dataset used was comprehensive, the effectiveness of UF$^3$ and similar models is limited when training data lacks representation for specific properties. Thus, ensuring diverse and representative training data is paramount for consistent and optimal model performance.

\subsection{Melting Point}

We then benchmark UF$^3$'s ability to predict more physically rich processes through melting-point predictions via LAMMPS. Here, melting points were estimated through a two-phase co-existence method~\cite{morris1994melting}, where the free energies of both liquid and solid phases are calculated within a single simulation box, from which the melting point can be accurately evaluated at the solid/liquid interface. This avoids hysteresis effects near the first-order transition, where transition temperatures can be significantly underestimated in simulated crystals. The coexistence simulations used supercells of 22{,}308 atoms (Ni, Cu), 11{,}154 atoms (Li, Mo), and 44{,}616 atoms (Si, Ge), built as $33\times13\times13$ replicas of the conventional cell with the long axis normal to the solid--liquid interface. After NPT equilibration and melting/requenching of one half of the box, the production stage evolves the two-phase system for 1.0--1.5 million steps (0.5--1.1~ns) in the NPH ensemble, in which the temperature is an outcome rather than a set point and the plateau temperature provides the melting-point estimate. Across the five systems with a converged plateau, the coexistence temperature is stable to $\pm 8$--$13$~K and the enthalpy drift is $(1$--$10)\times 10^{-8}$~eV/atom per step, below 0.1~eV/atom over the full production run. Table~\ref{tab:melting_points} shows the melting points calculated via UF$^3$ potentials compared to experimental values. We observe close agreement for Ni, Cu, and Li, with deviations of $+6.0\%$, $+5.1\%$, and $+0.2\%$ from experiment, whereas the melting point is underestimated by 19.9\% for Si and by 41.7\% for Mo. Despite achieving relatively faithful energy and force predictions for Ge, we were unable to create a stable potential to capture melting dynamics.

The transferability of UF$^3$ is element-dependent and reflects the bonding character of each system. For the simple metals Ni, Cu, and Li, where delocalized metallic bonding has weak angular character, UF$^3$ reproduces experimental melting points within $\sim$6\%. For Mo, a BCC transition metal with d-electron-dominated angular bonding, the predicted melting point deviates substantially, consistent with the difficulty all MLIPs in Ong et al.~\cite{doi:10.1021/acs.jpca.9b08723} encounter with Mo's migration energetics and with the Cauchy-relation limitation of pair-dominated potentials discussed in~\cite{xie2023ultrafast,ziegenhain2009pair}. For the diamond-cubic semiconductors Si and Ge, melting is accompanied by a semiconductor-to-metal transition: the fourfold covalent network transforms into a metallic liquid with coordination $\approx 6.4$~\cite{stich1989bonding,kresse1994liquidGe}, so the effective interactions change character between the two phases --- a regime change that a fixed two- plus three-body expansion cannot represent simultaneously, and that transferable empirical potentials capture only through environment-dependent bond-order rescaling as in Tersoff~\cite{tersoff1988new} or parameterizations explicitly tuned to the liquid as in Stillinger--Weber~\cite{stillinger1985computer}. Ge additionally exhibits small polymorph energy differences that destabilize the fit, preventing us from obtaining a Ge potential stable in MD; consistent with this, its learned pair interaction develops a spurious repulsive barrier between the first and second coordination shells (Fig.~\ref{fig:v2all} in \ref{app:v2all}), larger in magnitude than its own binding minimum, which anticipates the observed instability. Part of the Si discrepancy is furthermore inherited from the DFT reference itself: ab initio melting studies give $T_m \approx 1350\pm100$~K (LDA)~\cite{sugino1995ab} and $1590\pm50$~K (GGA)~\cite{alfe2003si} against the experimental 1687~K, so relative to the GGA reference appropriate for our PBE training data the UF$^3$ underestimate is reduced from 20\% to approximately 15\%.

\begin{table}[htbp]
\centering
\caption{Calculated vs experimental melting points. All calculations were performed in LAMMPS with a time step of 0.5~fs (1.0~fs for Cu, for which the coexistence plateau is already converged at this value; timestep convergence was verified for each element). The calculated value and its uncertainty are the mean and standard deviation of the temperature over the coexistence plateau of the production run. The last column gives the relative deviation from experiment.}
\label{tab:melting_points}
\begin{tabular}{lccc}
\toprule
\textbf{Element} & \textbf{Calculated (K)} & \textbf{Experimental (K)} & \textbf{Deviation} \\
\midrule
Ni & $1832 \pm 10$ & 1728 & $+6.0\%$ \\
Cu & $1427 \pm 8$  & 1358 & $+5.1\%$ \\
Li & $455 \pm 12$  & 454  & $+0.2\%$ \\
Mo & $1689 \pm 13$ & 2896 & $-41.7\%$ \\
Si & $1352 \pm 6$  & 1687 & $-19.9\%$ \\
Ge & $-$ & $1211$ & $-$ \\
\bottomrule
\end{tabular}
\end{table}

\subsection{Phonons, Thermal Expansion, and Dynamical Stability}

\begin{figure}[htbp]
\centering
\includegraphics[width=0.95\textwidth]{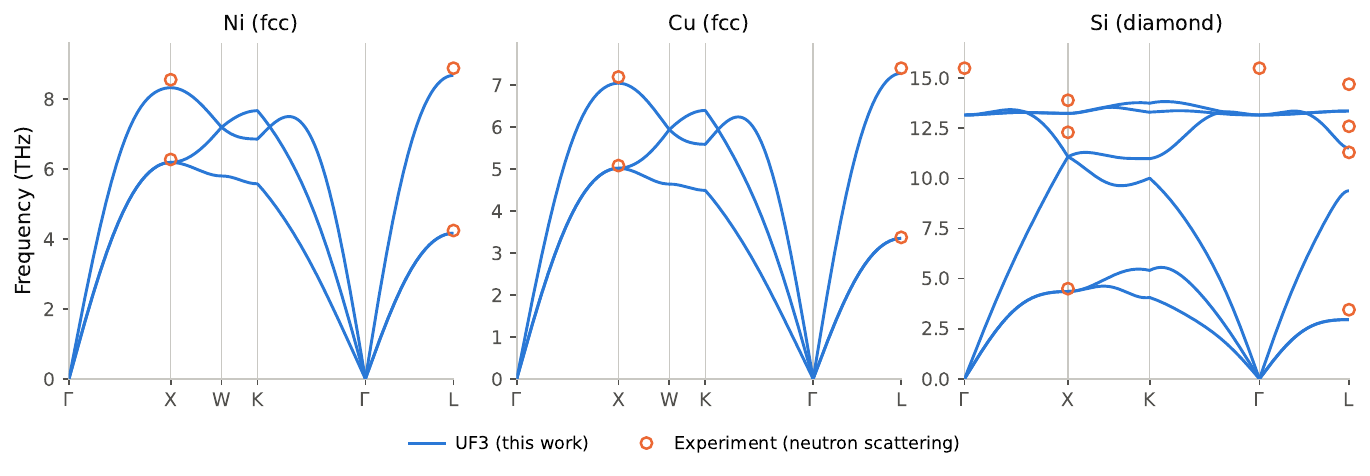}
\caption{Phonon dispersions of Ni, Cu, and Si computed with the final UF$^3$ potentials via the finite-displacement method. No imaginary modes appear anywhere on the high-symmetry paths. Open circles: room-temperature inelastic-neutron-scattering data at the high-symmetry points~\cite{birgeneau1964phonons,svensson1967copper,nilsson1972SiGe}.}
\label{fig:phonons}
\end{figure}

\begin{figure}[htbp]
\centering
\includegraphics[width=0.7\textwidth]{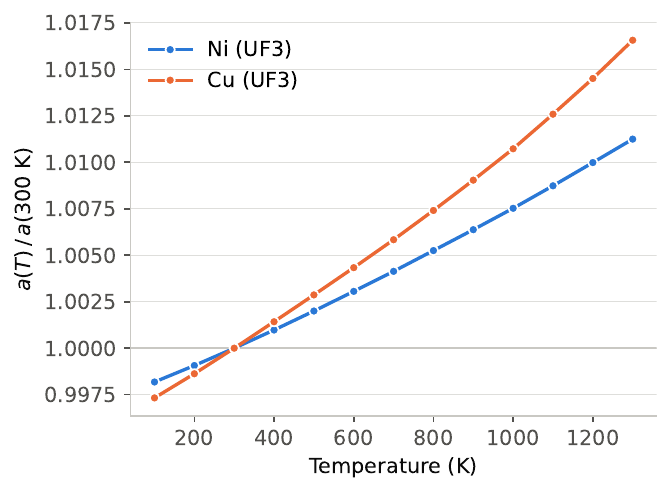}
\caption{Lattice parameter of Ni and Cu as a function of temperature from zero-pressure NPT simulations with the UF$^3$ potentials, normalized to $a(300~\mathrm{K})$. Experimental thermal-expansion data for comparison are tabulated in Ref.~\cite{touloukian1975thermal}.}
\label{fig:thermexp}
\end{figure}

To further probe whether the UF$^3$ potentials are practical for atomistic simulations beyond energy and force metrics, we computed three additional classes of validation properties. First, phonon dispersions for Ni, Cu, and Si obtained with the finite-displacement method (0.01~\AA\ displacements; 256--1{,}000-atom supercells, converged to below 0.001~THz against larger cells) exhibit no imaginary modes anywhere on the high-symmetry paths, confirming the dynamical stability of the predicted ground-state structures (Fig.~\ref{fig:phonons}). The Ni and Cu dispersions reproduce the measured fcc lattice dynamics --- including the TA/LA crossing between X and W and the K-point splittings --- with X- and L-point frequencies within 0.7--2.7\% of room-temperature inelastic-neutron-scattering data~\cite{birgeneau1964phonons,svensson1967copper}, without any vibrational information having entered the fits. For Si, the deviations are larger and uniformly soft~\cite{nilsson1972SiGe}: the transverse-acoustic branch at X lies within 3\% of experiment, but the softening grows to 14--17\% for the L-point acoustic modes and reaches 15\% for the zone-center optical frequency (13.2~THz against the measured $\approx$15.5~THz), consistent with the covalent-bonding limitations discussed above. Second, NPT simulations at zero pressure between 100 and 1300~K yield smooth lattice-parameter curves $a(T)$ with the expected mild anharmonic curvature and 300~K lattice constants within 0.2\% (Ni) and 0.9\% (Cu) of experiment (Fig.~\ref{fig:thermexp}); the mean linear thermal-expansion coefficients over 300--1000~K, $10.8\times10^{-6}$~K$^{-1}$ for Ni and $15.3\times10^{-6}$~K$^{-1}$ for Cu, underestimate the experimental values~\cite{touloukian1975thermal} by roughly 25--30\% and 15--20\%, respectively. This shortfall is expected, as no finite-temperature volumetric data constrained the fits, so the anharmonicity of the potential-energy surface is only weakly determined; for Ni, part of the measured expansion below the Curie temperature is moreover magnetic in origin and inaccessible to any non-magnetic interatomic potential. Third, 1~ns NVE molecular dynamics runs for all stable potentials ($\approx$4{,}000-atom cells, $\Delta t = 0.5$~fs, at 300~K and near $0.9\,T_m$) conserve energy with drifts between $10^{-10}$ and $4\times10^{-7}$~eV/(atom$\cdot$ns) --- an accumulated energy change below $10^{-6}$~eV/atom over the full nanosecond --- and 1~ns NPT trajectories at 300~K hold the cell volume constant to within 0.02\%/ns, with all final configurations remaining crystalline by the Bragg-peak structure factor. The only run that loses its crystal structure is Mo at $0.9\,T_m^{\rm exp} = 2606$~K, which lies 900~K above the potential's own melting point, where melting is the correct physical outcome (the Mo crystal is stable at $0.9$ of its UF$^3$ melting point); conversely, defect-free bulk Si remains crystalline even above its UF$^3$ melting point, a superheating effect that illustrates why the melting points above were obtained from two-phase coexistence rather than from heating a bulk crystal. Together with the elastic-property benchmarks of Fig.~\ref{fig:spider} and the long two-phase coexistence trajectories of the melting-point calculations, these tests support the practicality of UF$^3$ potentials for production molecular dynamics for the systems that pass the diagnostic screening of the learned interactions in Section~3.1 --- while Ge, which fails that screening, is correctly excluded.

\section{Discussion and Conclusion}
Our study demonstrates that optimizing the UF$^3$-MLIP for energy and forces results in impressive performance, despite its straightforward physically-inspired structure. Notably, our optimization focused primarily on minimizing energy and force errors, which led to the employment of minimal regularization. However, we propose that a slight increase in curvature regularization may yield a smoother potential. This adjustment could marginally reduce the model's accuracy on energy and forces in the testing data, yet potentially enhance its application in property prediction and dynamics simulations.

It is also crucial to acknowledge that the UF$^3$ potentials, as utilized in this study, were specifically tailored for energy and force predictions. Their effectiveness in predicting other properties was not considered during the optimization process. Moving forward, we aim to incorporate these relevant properties directly into the optimization loop. This approach is intended not only to enhance accuracy but also to yield more functionally optimal potentials. Emphasizing the practicality of MLIPs in molecular dynamics simulations is essential, as the potential that returns the smallest errors on energies and forces is not always the most applicable or physically relevant \cite{fu2022forces, 10.1063/5.0139611}. The Ge potential of this work is a concrete example: its test errors are competitive, yet its learned interactions contain a spurious inter-shell barrier and its dynamics are unstable. We therefore advocate a three-stage validation pipeline for spline-based MLIPs --- energy/force accuracy, property-level validation (elastic constants, phonons, thermal expansion, melting), and direct inspection of the learned interactions --- in which the interpretability of UF$^3$ plays an essential role.

A notable result of this study is the element-dependent transferability of UF$^3$: potentials fitted without any solid--liquid interface configurations or explicit melting information reproduce experimental melting points within $\sim$6\% for the simple metals Ni, Cu, and Li, but transferability degrades systematically with increasing bonding complexity. The substantial deviation for Mo mirrors the difficulty all MLIPs in the Ong et al.~\cite{doi:10.1021/acs.jpca.9b08723} benchmark encounter with d-electron systems, while the diamond-cubic semiconductors Si and Ge undergo a change of bonding character upon melting~\cite{stich1989bonding,kresse1994liquidGe} that lies beyond UF$^3$'s fixed three-body truncation~\cite{stillinger1985computer,tersoff1988new}. Combined with UF$^3$'s inherent interpretability through direct visualization of its spline-based interactions --- which anticipates precisely these limits --- these results position UF$^3$ as a transparent framework whose physical scope is now better delineated.

Our subsequent steps include evaluating UF$^3$'s performance in MD simulations, especially for targeted applications. Additionally, we emphasize the need for tailored human oversight in MLIP training. For specific uses like elasticity calculations, training data should predominantly consist of bulk structures, which is feasible given UF$^3$'s reduced computational needs and effective training with less data. In conclusion, we aim to refine UF$^3$ optimizations to balance accuracy with practical utility in molecular dynamics simulations, ensuring the developed potentials are both precise and widely applicable.

\section*{Acknowledgments}
We extend our sincere thanks to Stephen R. Xie, Ajinkya Hire, and Jason Gibson for their invaluable discussions and insights that significantly contributed to this research. We are also grateful to the University of Florida for providing computational resources through the HiPerGator computing facility. This work was supported in part by the National Science Foundation, Division of Materials Research under Award No.~NSF-DMR-2118718 and by the Center for Molecular Magnetic Quantum Materials, an Energy Frontier Research Center funded by the U.S. Department of Energy, Office of Science, Basic Energy Sciences under Award No.~DE-SC0019330.

\section*{Data availability}
The data used in this work for training the potentials were obtained from Ref.~\cite{doi:10.1021/acs.jpca.9b08723} and are available at \url{https://github.com/materialyzeai/mlearn/tree/master/data}. The UF$^3$ code is available at \url{https://github.com/uf3/uf3}. The final fitted UF$^3$ potential files in LAMMPS-ready format and the fitting scripts used in this work are provided with the revised submission and will be deposited in a public repository upon publication.

\appendix

\section{Learned pair interactions for all six elements}
\label{app:v2all}

\begin{figure}[htbp]
    \centering
    \includegraphics[width=\textwidth]{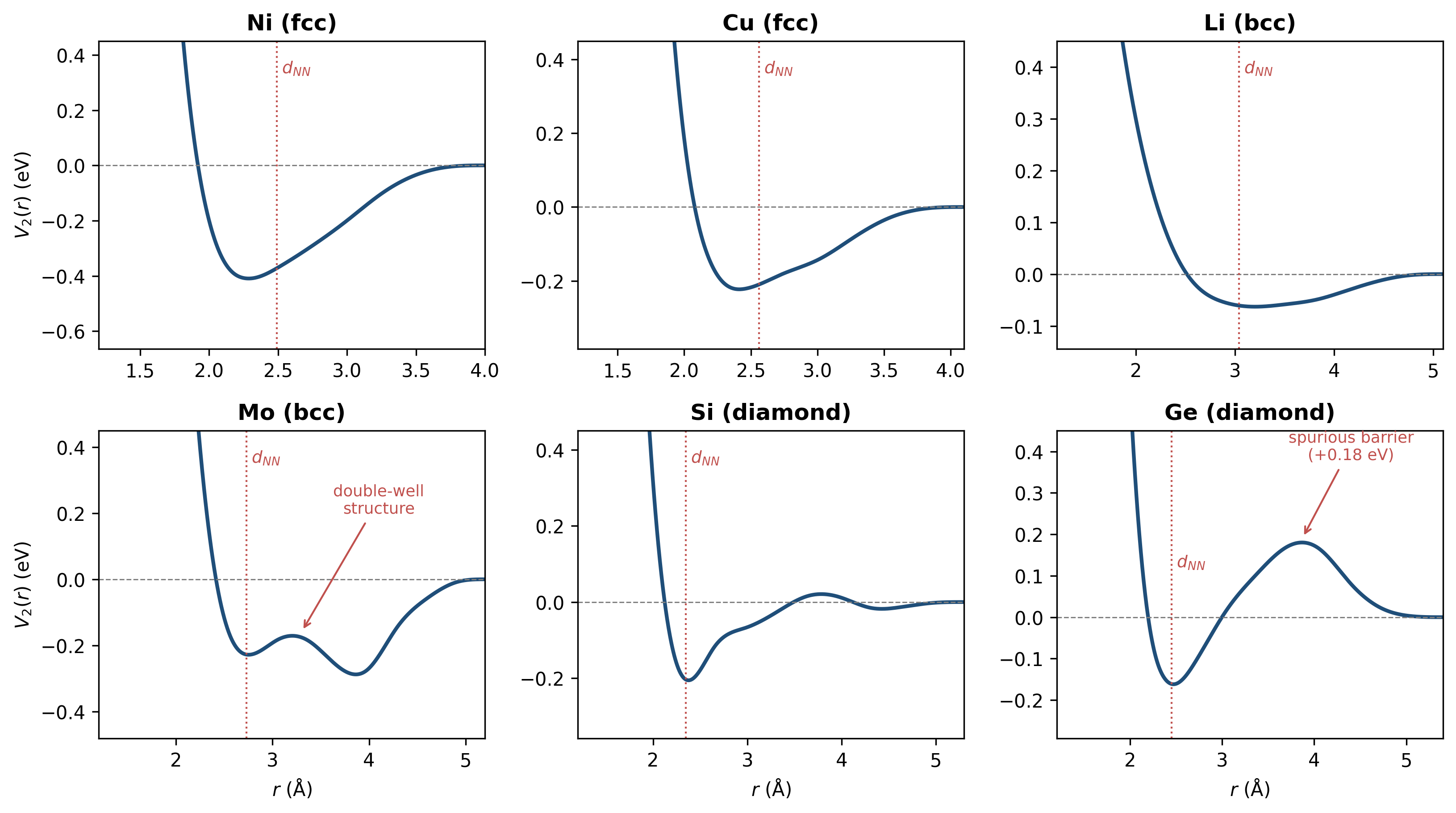}
    \caption{Learned UF$^3$ two-body terms $V_2(r)$ for all six elements. Dotted vertical lines mark the equilibrium nearest-neighbor distances $d_{NN}$. Ni, Cu, and Li display smooth, single-minimum pair interactions; Mo develops a double-well structure associated with the first- and second-neighbor shells of the bcc lattice; Ge exhibits a spurious repulsive barrier of $+0.18$~eV between its coordination shells --- nine times larger than the corresponding feature in Si ($+0.02$~eV) and larger in magnitude than its own binding minimum ($-0.16$~eV) --- which anticipates the instability of its melting simulations.}
    \label{fig:v2all}
\end{figure}

Figure~\ref{fig:v2all} shows the learned two-body terms $V_2(r)$ for all six elements, evaluated directly from the final fitted models. These curves underpin the diagnostic screening discussed in Section~3.1 and the analysis of the element-dependent melting-point transferability in Section~3.3.

\bibliographystyle{elsarticle-num}
\bibliography{reference}

\end{document}